\documentclass[pra,preprint,superscriptaddress,amssymb]{revtex4}
\usepackage{graphicx}
\usepackage{bm}
\usepackage{amsmath,amssymb,fancybox}
\usepackage{color}

\begin{document}
\title{Quantum Arthur-Merlin with single-qubit measurements} 
\author{Tomoyuki Morimae}
\email{morimae@gunma-u.ac.jp}
\affiliation{ASRLD Unit, Gunma University, 1-5-1 Tenjin-cho Kiryu-shi
Gunma-ken, 376-0052, Japan}

\begin{abstract}
We show that the class QAM does not change even if
the verifier's ability is restricted to only single-qubit measurements.
To show the result, we use the idea of
the measurement-based quantum computing:
the verifier, who can do only single-qubit measurements, 
can test the graph state sent from the prover
and use it for his measurement-based quantum computing.
We also introduce a new QMA-complete problem related
to the stabilizer test.
\end{abstract}
\maketitle  

\section{Introduction}
Measurement-based quantum computing~\cite{MBQC}
is a model of quantum computing where a universal quantum
computing is realized by preparing a highly-entangled quantum
many-qubit state, so called a resource state, and measuring
each qubit adaptively.
Mathematically it is equivalent to the standard quantum circuit model,
but the clear separation between the resource preparation phase
(i.e., preparation of a resource state with entangling gates) 
and the resource consumption
phase (i.e., single-qubit measurements) has enabled
plenty of new results not only in quantum computing~\cite{RHG}
but also in quantum cryptography~\cite{BFK,FK,measuringAlice,HayashiMorimae} 
and condensed matter physics~\cite{BrennenMiyake}.

The measurement-based quantum computing has
turned out to be useful also in quantum computational complexity.
In Ref.~\cite{Matt}, a quantum 
multiprover interactive proof system with a classical verifier
that contains BQP was constructed. Furthermore, 
in Ref.~\cite{MorimaeNagajSchuch} it was shown 
that the class QMA does not change
even if the verifier's ability is restricted to only single-qubit 
measurements.
The basic idea underlying these results is
the graph state verification:
the verifier, who is completely classical
or can do only single-qubit measurements,
can test the graph state generated by the prover(s)
and use it for his measurement-based quantum computing.

In this paper, we generalize the result of Ref.~\cite{MorimaeNagajSchuch}
to show that the class QAM does neither change
under the reduction of the verifier's ability to
single-qubit measurements.
Although QAM is somehow a generalized version of QMA,
the proof of Ref.~\cite{MorimaeNagajSchuch} cannot be directly
used for the QAM case, since the prover's strategy,
i.e., generating the correct graph state or trying to cheat the verifier,
can depend on the previous
message by the verifier. 
Furthermore, the local Hamiltonian technique used in 
Ref.~\cite{MorimaeNagajSchuch} can neither be used since there is no
known local Hamiltonian problem hard for QAM.
We nevertheless show that the reduction of the verifier's ability
is possible by using the graph state test.

The class QAM was introduced by Marriott and 
Watrous~\cite{MW}: 

{\bf Definition}:
For an instance $x\in \{0,1\}^*$,
we consider the following protocol, which we call a QAM system:
\begin{itemize}
\item[1.]
Arthur (the verifier) sends Merlin (the prover) a random $s(|x|)$-bit string 
$y\in\{0,1\}^{s(|x|)}$, where $s$ is a polynomial.
\item[2.]
Merlin sends Arthur an $m(|x|)$-qubit state $|\psi_{x,y}\rangle$, 
which can depend on $x$ and $y$,
where $m$ is a polynomial.
\item[3.]
Arthur applies a (uniformly generated) unitary $A_{x,y}$, 
which can depend on $x$ and $y$, on
$|\psi_{x,y}\rangle\otimes|+\rangle^{\otimes v(|x|)}$,
where $|+\rangle\equiv\frac{1}{\sqrt{2}}(|0\rangle+|1\rangle)$ 
and $v$ is a polynomial. (In the original definition,
ancilla qubits are initialized in $|0\rangle$. In this paper,
however, we initialize in $|+\rangle$, since it is more convenient
for the measurement-based quantum computing.)
Arthur measures the output
qubit. If the result is 1 (0), he accepts (rejects).
\end{itemize}
The acceptance probability $p_{acc}$ of the QAM system is defined by
\begin{eqnarray*}
p_{acc}\equiv \frac{1}{2^s}\sum_{y\in\{0,1\}^s}
\big\|\Pi_1A_{x,y}|\psi_{x,y}\rangle\otimes|+\rangle^{\otimes v}\big\|^2,
\end{eqnarray*}
where $\Pi_1\equiv|1\rangle\langle1|$ is the projection
on the output qubit.
A language $L$ is in QAM($a$,$b$) if and only if
there exists a QAM system such that
\begin{itemize}
\item
If $x\in L$, then there exists a set 
$\{|\psi_{x,y}\rangle\}_y$ 
of $m(|x|)$-qubit states
such that $p_{acc}\ge a$.
\item
If $x\notin L$, then for any set $\{|\psi_{x,y}\rangle\}_y$ 
of $m(|x|)$-qubit states, $p_{acc}\le b$.
\end{itemize}
As is shown in Ref.~\cite{MW}, the error bound $(a,b)$ can be
arbitrary: as long as $a-b\ge\frac{1}{poly(|x|)}$,
${\rm QAM}(a,b)\subseteq{\rm QAM}(1-2^{-r},2^{-r})$
for any polynomial $r$.

The main result of the present paper is
\begin{eqnarray*}
{\rm QAM}={\rm QAM}_{\rm single},
\end{eqnarray*}
where ${\rm QAM}_{\rm single}$ is equivalent to QAM except that
the verifier can do only single-qubit measurements.

In the discussion section,
we also introduce a new QMA-complete
problem that is related to the stabilizer test.

\section{Stabilizer test}
In this section, we explain the stabilizer test,
which will be used in this paper.
(The following stabilizer test is a generalized version
of that in Ref.~\cite{MorimaeNagajSchuch}: in 
Ref.~\cite{MorimaeNagajSchuch} only the graph state
stabilizers are considered, whereas in the following
the test is generalized to any stabilizers.)

The $N$-fold Pauli group is the set of $N$-fold tensor products
of Pauli operators:
\begin{eqnarray*}
\Big\{\{1,-1,i,-i\}\bigotimes_{j=1}^N\{I_j,X_j,Y_j,Z_j\}\Big\}.
\end{eqnarray*}
A stabilizer is an abelian subgroup of the $N$-fold Pauli group
not containing $-I^{\otimes N}$.

Let us consider an $N$-qubit state $\rho$ and
a set $g\equiv\{g_1,...,g_n\}$ of generators of a stabilizer.
The stabilizer test is the following test: 
\begin{itemize}
\item[1.]
Randomly generate an $n$-bit string
$k\equiv(k_1,...,k_n)\in\{0,1\}^n$. 
\item[2.]
Measure the operator
\begin{eqnarray*}
s_k\equiv\prod_{j=1}^ng_j^{k_j}.
\end{eqnarray*}
Note that this measurement can be done with single-qubit
measurements, since $s_k$ is a tensor product of Pauli operators:
\begin{eqnarray*}
\Big\{\{1,-1\}\bigotimes_{j=1}^N\{I_j,X_j,Y_j,Z_j\}\Big\}.
\end{eqnarray*}
(Note that the phase of $s_k$ cannot be $\pm i$, since if so
$s_k^2=-I^{\otimes N}$, which contradicts to the definition
of the stabilizer.)

\item[3.]
If the result is $+1$ ($-1$), the test passes (fails). 
\end{itemize}
The probability of passing the stabilizer test is
\begin{eqnarray*}
p_{pass}=\frac{1}{2^n}\sum_{k\in\{0,1\}^n}\mbox{Tr}
\Big(\frac{I^{\otimes N}+s_k}{2}\rho\Big).
\end{eqnarray*}
We can show that if the probability of passing
the test is high, which means 
$p_{pass}\ge1-\epsilon$ for an $\epsilon>0$, then
$\rho$ is ``close" to a certain stabilized state $\sigma$ in the sense of
\begin{eqnarray*}
\mbox{Tr}(M\sigma)(1-2\epsilon)-2\sqrt{2\epsilon}
\le
\mbox{Tr}(M\rho)\le\mbox{Tr}(M\sigma)+2\sqrt{2\epsilon}
\end{eqnarray*}
for any POVM element $M$.

In fact,
if $p_{pass}\ge1-\epsilon$, we obtain
\begin{eqnarray*}
\mbox{Tr}\Big(\prod_{j=1}^n\frac{I^{\otimes N}+g_j}{2}\rho\Big)\ge1-2\epsilon.
\end{eqnarray*}
Let
\begin{eqnarray*}
\Lambda\equiv
\prod_{j=1}^n\frac{I^{\otimes N}+g_j}{2}.
\end{eqnarray*}
Note that $\Lambda^\dagger=\Lambda$
and $\Lambda^2=\Lambda$,
and therefore $0\le \Lambda\le I^{\otimes N}$.
From the gentle measurement lemma~\cite{gentle,gentle2},
\begin{eqnarray}
\|\rho-\Lambda \rho\Lambda\|_1&\le&
2\sqrt{1-\mbox{Tr}(\Lambda\rho)}\nonumber\\
&\le&2\sqrt{1-(1-2\epsilon)}\nonumber\\
&=&2\sqrt{2\epsilon}.
\label{gentle_ineq}
\end{eqnarray}
Note that 
\begin{eqnarray*}
g_j\frac{\Lambda\rho\Lambda}{\mbox{Tr}(\Lambda\rho)}g_j
=
\frac{\Lambda\rho\Lambda}{\mbox{Tr}(\Lambda\rho)}
\end{eqnarray*}
for any $j$, and therefore, 
\begin{eqnarray*}
\sigma\equiv\frac{\Lambda\rho\Lambda}{\mbox{Tr}(\Lambda\rho)} 
\end{eqnarray*}
is a stabilized state.

From Eq.~(\ref{gentle_ineq}), we obtain for any POVM element $M$
\begin{eqnarray*}
\mbox{Tr}(M\rho)-\mbox{Tr}(M\Lambda\rho\Lambda)\le2\sqrt{2\epsilon},
\end{eqnarray*}
which means
\begin{eqnarray*}
\mbox{Tr}(M\rho)&\le&
\mbox{Tr}\Big(M\frac{\Lambda\rho\Lambda}
{\mbox{Tr}(\Lambda\rho)}\Big)
{\mbox{Tr}(\Lambda\rho)}
+2\sqrt{2\epsilon}\\
&\le&
\mbox{Tr}\Big(M\frac{\Lambda\rho\Lambda}
{\mbox{Tr}(\Lambda\rho)}\Big)
+2\sqrt{2\epsilon}.
\end{eqnarray*}
And, for any positive operator $M$,
\begin{eqnarray*}
\mbox{Tr}(M\Lambda\rho\Lambda)-\mbox{Tr}(M\rho)
\le2\sqrt{2\epsilon},
\end{eqnarray*}
which means
\begin{eqnarray*}
\mbox{Tr}(M\rho)
&\ge&
\mbox{Tr}\Big(M\frac{\Lambda\rho\Lambda}{\mbox{Tr}(\Lambda\rho)}\Big)
\mbox{Tr}(\Lambda\rho)
-2\sqrt{2\epsilon}\\
&\ge&
\mbox{Tr}\Big(M\frac{\Lambda\rho\Lambda}{\mbox{Tr}(\Lambda\rho)}\Big)
(1-2\epsilon)
-2\sqrt{2\epsilon}.
\end{eqnarray*}

\section{Proof of the result}
\begin{figure}[htbp]
\begin{center}
\includegraphics[width=0.4\textwidth]{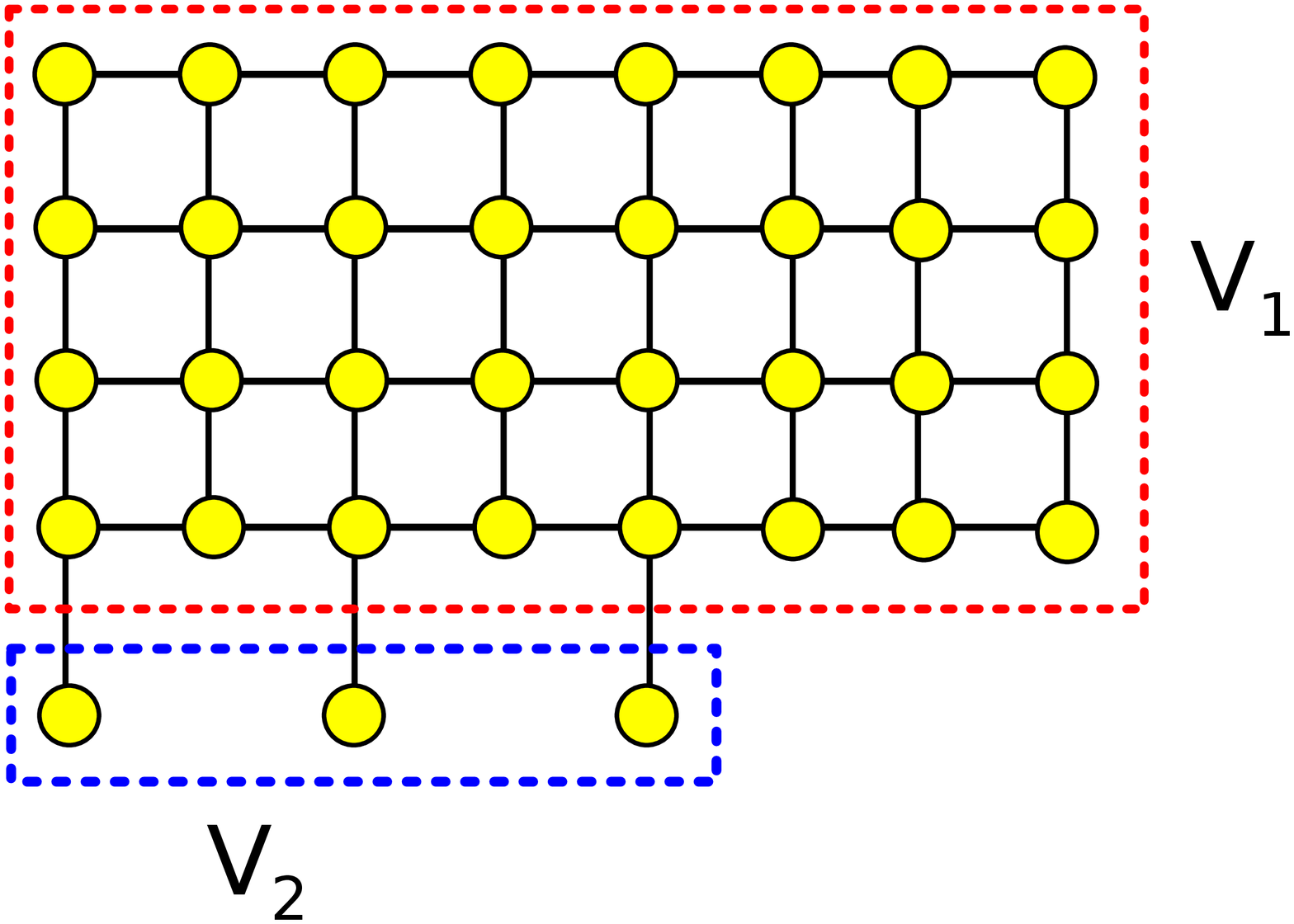}
\end{center}
\caption{
The system that Merlin sends to Arthur.
} 
\label{fig}
\end{figure}

Now let us show our main result,
${\rm QAM}_{\rm single}={\rm QAM}$.
The inclusion ${\rm QAM}\supseteq{\rm QAM}_{\rm single}$ is obvious.
We show the inverse
${\rm QAM}\subseteq{\rm QAM}_{\rm single}$.
Let us assume that a language $L$ is in QAM.
From the corresponding QAM system, we construct the 
following ${\rm QAM}_{\rm single}$ system:
\begin{itemize}
\item[1.]
Arthur sends Merlin a random $s(|x|)$-bit string 
$y\in\{0,1\}^{s(|x|)}$.
\item[2.]
Honest Merlin generates the state
\begin{eqnarray}
\big(\bigotimes_{e\in E_{connect}}CZ_e\big)
\big(|\psi_{x,y}\rangle\otimes |G\rangle\big), 
\label{honestMerlin}
\end{eqnarray}
where $|\psi_{x,y}\rangle$ is the $m$-qubit state 
(the original witness) on the subsystem $V_2$,
$|G\rangle$ is the $N$-qubit graph state on the subsystem $V_1$,
and $E_{connect}$ is the set of edges that connect $V_1$ and $V_2$
(see Fig.~\ref{fig}),
where $N=poly(|x|)$.
Malicious Merlin generates any $(m+N)$-qubit state $\rho_{x,y}$.
Merlin sends qubits of his $(m+N)$-qubit state one by one to Arthur. 
\item[3.]
With probability $q$, which is specified later,
Arthur does the measurement-based quantum computing on
qubits sent from Merlin.
If the computation accepts (rejects), Arthur accepts (rejects).
With probability $1-q$, Arthur does the 
following stabilizer test:
Arthur generates a random $N$-bit string 
$k\equiv(k_1,...,k_N)\in\{0,1\}^N$, and measures the operator
\begin{eqnarray*}
s_k\equiv\prod_{j\in V_1}g_j^{k_j},
\end{eqnarray*}
where 
\begin{eqnarray*}
g_j\equiv X_j\bigotimes_{i\in S_j}Z_i
\end{eqnarray*}
is the graph state stabilizer on $j$th qubit.
Here, $S_j$ is the set of the nearest-neighbour qubits of $j$th qubit.
If the measurement result is +1 ($-1$), Arthur accepts (rejects).
\end{itemize}

If $x\in L$, Merlin is honest, and therefore he generates the 
state of Eq.~(\ref{honestMerlin}).
In this case, Arthur can do the correct computation, i.e.,
$A_{x,y}$, if he chooses computation, and passes the stabilizer
test with probability 1 if he chooses the test.
Therefore, the total acceptance probability $p_{acc}^{x\in L}$ is
\begin{eqnarray*}
p_{acc}^{x\in L}&=&\frac{1}{2^s}\sum_{y\in\{0,1\}^s}
\Big[q\big\|\Pi_1A_{x,y}|\psi_{x,y}\rangle
\otimes|+\rangle^{\otimes v}\big\|^2+(1-q)\times 1
\Big]\\
&\ge&qa+(1-q)\equiv\alpha.
\end{eqnarray*}

If $x\notin L$, Merlin is malicious, and he can send any state. 
Let $Y_1\subseteq\{0,1\}^s$ be the set of $y$ such that
$\rho_{x,y}$ passes the stabilizer test with probability $\ge 1-\epsilon$,
where $\epsilon=\frac{1}{128|x|^2}$.
We also define $Y_2\equiv\{0,1\}^s-Y_1$. Note that
$|Y_1|+|Y_2|=2^s$.
Let $p_{MBQC}^y$ be the acceptance probability 
when Arthur does the measurement-based quantum computing
on $\rho_{x,y}$, and let $p_{test}^y$ be the acceptance
probability when Arthur does the stabilizer test on $\rho_{x,y}$.
Then, the total acceptance probability $p_{acc}^{x\notin L}$ is
\begin{eqnarray*}
p_{acc}^{x\notin L}&=&\frac{1}{2^s}\sum_{y\in\{0,1\}^s}
\Big[
q p_{MBQC}^y+(1-q)p_{test}^y
\Big]\\
&=&\frac{1}{2^s}\sum_{y\in Y_1}
\Big[
q p_{MBQC}^y+(1-q)p_{test}^y
\Big]
+\frac{1}{2^s}\sum_{y\in Y_2}
\Big[
q p_{MBQC}^y+(1-q)p_{test}^y
\Big]\\
&<&\frac{1}{2^s}\sum_{y\in Y_1}
\Big[
q\big(\|\Pi_1A_{x,y}|\psi_{x,y}\rangle\otimes|+\rangle^{\otimes v}\|^2+
\delta\big)+(1-q)
\Big]
+\frac{1}{2^s}\sum_{y\in Y_2}
\Big[
q +(1-q)(1-\epsilon)
\Big]\\
&\le&
qb
+\frac{(q\delta+1-q)|Y_1|}{2^s}
+\frac{
(q +(1-q)(1-\epsilon))
|Y_2|}{2^s}\\
&=&
qb
+\frac{(q\delta+1-q)|Y_1|}{2^s}
+\frac{(q\delta+1-q)|Y_2|}{2^s}
-\frac{(q\delta+1-q)|Y_2|}{2^s}
+\frac{
(q +(1-q)(1-\epsilon))
|Y_2|}{2^s}\\
&=&
qb
+q\delta+1-q
+\frac{
(q -q\delta+q\epsilon-\epsilon)
|Y_2|}{2^s},
\end{eqnarray*}
where $\delta=2\sqrt{2\epsilon}$.
If we take
\begin{eqnarray*}
q=\frac{\epsilon}{1+\epsilon-\delta},
\end{eqnarray*}
we obtain
\begin{eqnarray*}
q-q\delta+q\epsilon-\epsilon=0.
\end{eqnarray*}
Therefore,
\begin{eqnarray*}
p_{acc}^{x\notin L}&\le&qb+
q\delta+1-q\\
&\equiv&\beta.
\end{eqnarray*}
Then, if $a=\frac{2}{3}$ and $b=\frac{1}{3}$,
\begin{eqnarray*}
\alpha-\beta&\ge& q(a-b-\delta)\\
&=&
\frac{(a-b-\delta)\epsilon}{1+\epsilon-\delta}\\
&\ge&\frac{1}{12\times129|x|^2}.
\end{eqnarray*}
Therefore, $L$ is in ${\rm QAM}_{single}(\alpha,\beta)$
with $\alpha-\beta\ge\frac{1}{poly(|x|)}$.

\section{Discussion: a QMA-complete problem}
In this paper, we have shown that by using the stabilizer test,
the verifier's ability of QAM
can be reduced to the single-qubit measurements.

In this section, we point out that the stabilizer test
also gives a new QMA complete problem:
\begin{itemize}
\item
Input: the set $g\equiv(g_1,...,g_k)$ of generators of
an $n$-qubit stabilizer, an $n$-qubit 
POVM element $M$, and $0\le b<a\le 1$ such that
$a-b\ge1/poly(n)$.
\item
Task:
decide whether $h_{Stab(g)}(M)$ is $\ge a$ or $\le b$.
\end{itemize}
Here, 
\begin{eqnarray*}
h_{Stab(g)}(M)\equiv\max_\rho\{\mbox{Tr}(M\rho):\rho\in Stab(g)\},
\end{eqnarray*}
and
$Stab(g)$ be the set of $n$-qubit states that are stabilized by $g$,
i.e., states such that $g_i|\psi\rangle=|\psi\rangle$ for all $i=1,...,k$.

Note that if we replace $Stab(g)$ in the above definition
with $Sep(d,d)$, which is the set of separable states,
it is known that to calculate it up to 1/poly($d$) accuracy
is NP-hard~\cite{Gurvits},
and to estimate it to a constant additive error
is QMA(2)-complete~\cite{HarrowMontanaro}.

We first show the above problem is QMA-hard.
Let $A$ be a problem in QMA, and $|\psi\rangle$
be a witness of a yes instance.
We take $g=\{g_j\}_{j\in V_1}$ as the graph state stabilizers,
and $M$ be the POVM element corresponding
to the acceptance of the measurement-based quantum
computing that simulates the verification circuit.
Then,
if $x\in A_{yes}$, 
\begin{eqnarray*}
h_{Stab(g)}(M)\ge\mbox{Tr}(M G_\psi)\ge a,
\end{eqnarray*}
and if $x\in A_{no}$,
\begin{eqnarray*}
h_{Stab(g)}(M)=\max_\xi\{\mbox{Tr}(M G_\xi)\}\le b.
\end{eqnarray*}
Here, $G_\psi$ and $G_\xi$ are states constructed 
by connecting $\psi$ and $\xi$ to the graph state by $CZ$ gates
as Eq.~(\ref{honestMerlin}).

We next show the problem is in QMA.
In the yes case, the prover sends $\rho$
that maximizes $h_{Stab(g)}(M)$ to the verifier.
With probability $q$,
which is specified later,
the verifier does a POVM measurement that contains
$M$. If $M$ is realized, he accepts.
With probability $1-q$, the verifier
does the stabilizer test, and if passes, he accepts.
The acceptance probability is 
\begin{eqnarray*}
p_{acc}&=& q\mbox{Tr}(M\rho)+(1-q)\\
&\ge& qa+(1-q)\equiv\alpha.
\end{eqnarray*}
In the no case, the prover sends any state $\rho$.
Let $p_{pass}$ be the probability of passing the stabilizer
test, and let us define $\epsilon=\frac{(a-b)^2}{32}$. 
If $p_{pass}< 1-\epsilon$,
The acceptance probability is
\begin{eqnarray*}
p_{acc}&=&q\mbox{Tr}(M\rho)+(1-q)p_{pass}\\
&\le&q+(1-q)(1-\epsilon)\equiv\beta_1.
\end{eqnarray*}

On the other hand, if $p_{pass}\ge1-\epsilon$,
\begin{eqnarray*}
\mbox{Tr}(M\rho)&\le&
\mbox{Tr}\Big[M\frac{\Lambda\rho\Lambda}{\mbox{Tr}(\Lambda\rho)}\Big]
+\delta\\
&\le&b+\delta,
\end{eqnarray*}
and therefore
\begin{eqnarray*}
p_{acc}&=&q\mbox{Tr}(M\rho)+(1-q)p_{pass}\\
&\le&q(b+\delta)+(1-q)\equiv\beta_2,
\end{eqnarray*}
where $\delta=2\sqrt{2\epsilon}$.
Let us define
\begin{eqnarray*}
\Delta_1\equiv\alpha-\beta_1,\\
\Delta_2\equiv\alpha-\beta_2.
\end{eqnarray*}
Let us take $q$ as the value 
\begin{eqnarray*}
q^*\equiv\frac{\epsilon}{1+\epsilon-b-\delta},
\end{eqnarray*}
which satisfies
$\Delta_1(q^*)=\Delta_2(q^*)$.
Then,
\begin{eqnarray*}
\Delta_2(q^*)&=&\frac{\epsilon(a-b-\delta)}{1+\epsilon-b-\delta}\\
&\ge&\frac{(a-b)^3}{32\times4}\\
&\ge&\frac{1}{poly(n)}.
\end{eqnarray*}
Therefore the problem is in QMA.

\acknowledgements
TM is supported by the
Grant-in-Aid for Scientific Research on Innovative Areas
No.15H00850 of MEXT Japan, and the Grant-in-Aid
for Young Scientists (B) No.26730003 of JSPS.

\end{document}